\documentclass[10pt,conference]{IEEEtran}

\usepackage{amsmath}
\usepackage{amssymb}
\usepackage{booktabs}
\usepackage{cite}
\usepackage{graphicx}
\usepackage{xcolor}
\usepackage{array}
\usepackage{listings}
\usepackage[T1]{fontenc}
\usepackage{url}
\usepackage{algorithm}
\usepackage{algpseudocode}
\usepackage{xspace}

\newcommand{\tool}{SkillBridge\xspace}
\newcommand{\linux}{Linux\xspace}
\newcommand{\macos}{macOS\xspace}
\newcommand{\windows}{Windows\xspace}
\newcolumntype{L}[1]{>{\raggedright\arraybackslash}p{#1}}
\definecolor{diffred}{HTML}{B42318}
\newif\ifreviewcopy
\reviewcopyfalse
\newcommand{\rev}[1]{\ifreviewcopy\textcolor{diffred}{#1}\else#1\fi}
\definecolor{diffgreen}{HTML}{067647}
\definecolor{listingbg}{HTML}{F7F7F7}
\definecolor{listingrule}{HTML}{B8B8B8}
\lstdefinestyle{repairdiff}{
  basicstyle=\ttfamily\scriptsize,
  backgroundcolor=\color{listingbg},
  frame=single,
  framerule=0.3pt,
  rulecolor=\color{listingrule},
  numbers=left,
  numberstyle=\tiny\color{gray},
  numbersep=5pt,
  xleftmargin=1.1em,
  framexleftmargin=1.1em,
  columns=fullflexible,
  keepspaces=true,
  showstringspaces=false,
  breaklines=true,
  aboveskip=2pt,
  belowskip=2pt,
  morecomment=[f][\color{diffgreen}][0]{+},
  morecomment=[f][\color{diffred}][0]{-}
}

\begin{document}
\bstctlcite{IEEEtranBSTControl}

\title{SkillBridge: Cost-Aware Diagnosis and Repair of Cross-OS Compatibility Defects in Agent Skills}

\author{%
  \IEEEauthorblockN{Anjie Xu\textsuperscript{1,2,3},
  Zixing Wang\textsuperscript{4}, Yifeng Cai\textsuperscript{1,2},
  \rev{Leye Wang\textsuperscript{1,2,*}}}
  \IEEEauthorblockA{%
    \textsuperscript{1}Key Lab of High Confidence Software Technologies
      (Peking University), Ministry of Education, China\\
    \textsuperscript{2}School of Computer Science, Peking University, Beijing, China\\
    \textsuperscript{3}Zhongguancun Academy, Beijing, China\\
    \textsuperscript{4}School of Computer Science, Northwestern Polytechnical
      University, Xi'an, China\\
    anjiexu@stu.pku.edu.cn, leyewang@pku.edu.cn\\
    \textsuperscript{*}Corresponding author.}}

\maketitle

\begin{abstract}
LLM agents increasingly reuse executable capabilities through skills, but a
skill that works on one OS may fail on another because OS assumptions are spread
across guide text, scripts, metadata, install contracts, and local entrypoints.
Diagnosing such failures by running every skill on every OS is expensive: many
executions spend model tokens, runner time, and review effort before revealing
whether the failure is truly OS-related. We present \tool, a contract-based
workflow with three stages: static analysis, dynamic validation, and repair.
\tool first screens 3,270 skill-target-OS cases, covering 1,090 skills across
\linux, \macos, and \windows, to find OS-sensitive contract evidence. It then
runs dynamic tasks and root-cause review on selected cases to confirm whether a
failure is caused by an OS assumption, and repairs the contract field that
caused the failure, such as guide text, package scripts, dependencies, or
platform declarations. \rev{Repair raises the reviewer-judged target-task pass
rate from 30.6\% to 38.4\%, a gain of 7.8 percentage points.} A
retrospective \linux/\macos budget analysis further shows the efficiency
effect: \tool reaches 38 confirmed OS issues with 185 dynamic validations,
where uniform random selection would require about 865 validations in
expectation.
\end{abstract}


\section{Introduction}

Agents such as Codex, Claude Code, OpenCode, and OpenClaw are becoming
practical interfaces for software work and other tool-mediated tasks: they read
files, call tools, run terminal commands, and carry state across execution
loops~\cite{anthropicHarnesses2025,openaiCodex2026,
anthropicClaudeCodeOverview2026,opencode2026,openclaw2026}. To make these
capabilities reusable, several agent systems expose \emph{skills} as a packaging
unit for task-specific knowledge, scripts, dependencies, and execution
instructions~\cite{openaiCodexSkills2026,anthropicClaudeCodeSkills2026,
openclawClawHub2026}.

A skill can contain a \texttt{SKILL.md} guide, examples, helper scripts,
metadata, install instructions, referenced commands, local entrypoints, external
dependencies, and assumptions about the operating system where the harness runs.
This makes a skill a software-engineering artifact, not just a natural-language
prompt. Users expect the same skill to work on \linux servers, \macos laptops,
and \windows desktops, but many skills are authored and tested on only one OS.

These assumptions surface as task failures that are difficult to attribute. The
same skill may pass on one OS and fail on another, yet the difference may come
from an OS assumption, missing credentials, an unavailable service, a benchmark
fixture, or a non-OS agent planning error. The OS-specific surface is often
hidden inside the package: a guide may ask the agent to run
\texttt{./script.sh}, a package may omit the referenced entrypoint, an install
contract may only mention Homebrew or \texttt{apt}, or a script may depend on
AppleScript~\cite{appleScriptGuide}, Windows COM~\cite{microsoftCOM}, POSIX
paths~\cite{posix2024}, \texttt{/proc}~\cite{linuxProc}, or Windows naming
rules and path limits~\cite{microsoftWindowsNaming,microsoftMaxPath}. By the
time the agent discovers the mismatch dynamically, it has already spent model
tokens, runner time, and review effort.

Existing software portability work provides a foundation, but it does not
directly match this mixed artifact. Traditional analyses focus on source code,
APIs, system calls, tests, or dependency declarations~\cite{mooney1990,
iso25010,silva2026portability}. Agent benchmarks report task-level pass/fail
outcomes, but a failed task result is not an OS root cause. Automated repair
work typically patches code, while skill repair may require coordinated changes
to guide text, scripts, metadata, install contracts, and support declarations.
The missing abstraction is a unit that connects what the agent is told to read,
install, and execute with the OS assumption that explains a failure and the
contract field that a repair must change.

\tool diagnoses cross-OS skill compatibility as a contract problem. A dynamic
failure is only a symptom; it becomes an OS issue when review ties that symptom
to an operating-system assumption in the skill contract. This formulation also
turns efficiency into a selection problem. Static evidence proposes which
skill-OS pairs should receive dynamic validation, review decides the final
label, and dynamic execution is reserved for pairs that need runtime evidence.

We present \tool, a workflow for cost-aware cross-OS skill diagnosis and repair.
\tool builds a compact contract, extracts and interprets OS-sensitive evidence
for each target OS, allocates dynamic budget to selected pairs, and reviews root
causes. Repairs edit the contract component that caused the defect, such as
command branches, wrappers, install instructions, or support declarations.

Our evaluation covers 1,090 skills across three OSes, or 3,270
skill-target-OS cases. Static analysis screens these cases for OS-sensitive
contract evidence. A separate
\linux/\macos budget analysis shows that static-guided selection reaches the
same expected issue yield with 185 validations rather than about 865 under
uniform random selection.

The paper makes three contributions:
\begin{itemize}
  \setlength{\itemsep}{0pt}
  \setlength{\parsep}{0pt}
  \setlength{\topsep}{2pt}
  \item We define the cross-OS compatibility problem for agent skills: when the
  same skill task runs reliably on some OSes but persistently fails on a target
  OS, it is a candidate OS issue; review confirms the issue only when the
  failure is caused by an OS-specific assumption in the skill contract.

  \item We present \tool, which uses static triage to focus dynamic validation
  and a software engineering-inspired multi-agent workflow to triage, patch, and
  review repairs.

  \item We evaluate \tool on 1,090 public skills across \linux, \macos, and
  \windows. \rev{We open-source our experimental code and data through the
  public SkillBridge repository.}\footnote{\rev{\url{https://github.com/TankTechnology/SkillBridge}}}
\end{itemize}

\section{Problem Definition}
\label{sec:problem}

We study the following observable symptom: the same skill is expected to run on
$O=\{\mathsf{Linux},\mathsf{macOS},\mathsf{Windows}\}$, but it passes on some
OSes and fails on others. This symptom is insufficient by itself. We count an
OS issue only when the failure can be traced to the skill's stated or implicit
OS assumptions.

The central object is the skill contract: everything the skill exposes to the
agent as the way to understand, install, and execute the capability. We
write this contract as:

\[
  C(s) = \langle G_s, P_s, M_s, D_s, A_s \rangle .
\]

Here, $G_s$ is guide text and examples; $P_s$ is the package surface, including
files, scripts, wrappers, and referenced entrypoints; $M_s$ is metadata and
supported-platform declarations; $D_s$ is the dependency and installation
contract; and $A_s$ is the set of OS-sensitive assumptions, including shell
semantics, commands, paths, permissions, GUI/session requirements, native
applications, APIs, and platform guards. Assumptions in $A_s$ may be stated in
$G_s$, implemented in $P_s$, advertised in $M_s$, or required by $D_s$; the
separate component lets review name the OS assumption while repair edits the
field that exposes it. We merge files and referenced entrypoints into $P_s$
because the agent executes them as one package surface. These assumptions are
grounded in standard and vendor-documented OS surfaces, including POSIX shell
and utility behavior~\cite{posix2024}, Windows path
namespaces~\cite{microsoftWindowsNaming}, AppleScript~\cite{appleScriptGuide},
COM~\cite{microsoftCOM}, and Linux procfs~\cite{linuxProc}.

For each skill-OS pair $(s,o)$ that receives validation, \tool preserves one
evidence chain:

\[
\begin{aligned}
C(s) &\xrightarrow{\mathit{static}} (e(s,o),h(s,o)),\\
(e(s,o),h(s,o)) &\xrightarrow{\mathit{dynamic}} \tau(s,o),\\
\tau(s,o) &\xrightarrow{\mathit{review}} y(s,o),\\
y(s,o) &\xrightarrow{\mathit{repair}} r .
\end{aligned}
\]

Here, $e(s,o)$ is the static evidence extracted from $C(s)$ for target OS $o$;
$h(s,o)$ is the static hypothesis used to select the pair for dynamic checking;
$\tau(s,o)$ is the dynamic trace when the pair is run; $y(s,o)$ is the reviewed
label; and $r$ is an optional repair action. We use
$y(s,o)=\mathsf{confirmed}$ when review ties a cross-OS symptom to an OS
assumption in $C(s)$, $y(s,o)=\mathsf{excluded}$ when the symptom comes from a
non-OS cause, and $y(s,o)=\mathsf{uncertain}$ when the evidence is insufficient.

The formulation separates a dynamic symptom from a compatibility defect. We
write a reviewed OS issue as:

\begin{multline*}
\mathsf{OSIssue}(s,o) \Longleftrightarrow
  \mathsf{symptom}(\tau(s,o)) \wedge \exists a \in A_s.\\
  \mathsf{violates}(a,o) \wedge
  \mathsf{explains}(a,\tau(s,o)).
\end{multline*}

Here, $a$ is an OS-sensitive assumption in the skill contract.
$\mathsf{violates}$ means that the assumption is incompatible with the target
OS, and $\mathsf{explains}$ means that review can connect the dynamic symptom to
that violated assumption. Thus, a cross-OS pass/fail difference is evidence, but
it is not by itself an OS-issue label.

The cost-aware validation problem, for a skill set $S$, is to choose which
skill-OS pairs receive dynamic traces under a budget:

\[
\begin{aligned}
Q_B^* = \arg\max_{Q \subseteq S\times O} \quad&
  \sum_{(s,o)\in Q} \mathbf{1}[\mathsf{OSIssue}(s,o)]\\
\text{s.t.}\quad&
  \sum_{(s,o)\in Q} \mathsf{cost}(s,o) \le B .
\end{aligned}
\]

Because $\mathsf{OSIssue}(s,o)$ is unknown before dynamic execution and review,
\tool uses static evidence $e(s,o)$ and hypothesis $h(s,o)$ as a proxy for this
selection problem.

The same chain defines all three evaluation questions. RQ1 asks whether the
pre-execution hypothesis $h(s,o)$ can recover labels that root-cause review
later assigns as $y(s,o)$. RQ2 starts from confirmed labels and asks how a
repair $r:C(s)\rightarrow C'(s)$ changes aggregate reviewer-judged target-OS
replay and the per-pair outcome distribution. RQ3 asks whether the workflow
matters beyond that endpoint, comparing a direct-agent ablation and budgeted
selection policies that choose which $(s,o)$ pairs receive dynamic traces
$\tau(s,o)$.

A confirmed issue must be explained by a violated OS assumption in the
contract. For example, a \windows failure is an OS issue if the only executable
path in $P_s$ is a POSIX \texttt{.sh} script~\cite{posix2024}; a \linux failure
is an OS issue if $A_s$ depends on an Apple-only application~\cite{
appleScriptGuide}; a \macos failure is an OS issue if $G_s$ uses GNU-only
command flags without a BSD fallback~\cite{gnuCoreutils,freebsdDate}. Each
proposed repair receives a scoped replay label: the same target task is rerun
after the contract change, and the review stage assigns the replay judgment.
The label supports target-OS repair effectiveness, while unreplayed OSes remain
outside the claim.

\section{Method: \tool}
\label{sec:method}

The formulation imposes two requirements on \tool. First, \tool must preserve
the evidence chain from a contract assumption to a runtime symptom and then to a
reviewed label, which instantiates $\mathsf{OSIssue}$. Second, \tool must
allocate dynamic budget to pairs where low-cost static evidence predicts
OS-sensitive behavior, which approximates $Q_B^*$. \tool implements these
requirements in four stages: extract contract evidence from $C(s)$, interpret
that evidence for each target OS, validate selected pairs dynamically, and
convert reviewed root causes into contract-preserving repairs.

The design adapts a common software-engineering pattern to skills. Low-cost
structural evidence guides expensive execution~\cite{godefroid2005dart};
repair studies report both success and cost~\cite{
legoues2012genprog,marginean2019sapfix}; and inspection or review turns
symptoms into auditable labels~\cite{fagan1976inspection,sadowski2018review}.
\tool applies this pattern to a mixed text-code-package artifact by treating
guide text, package files, metadata, dependencies, and OS assumptions as one
contract rather than as separate artifacts.

\begin{figure*}[t]
  \centering
  \includegraphics[width=0.92\textwidth]{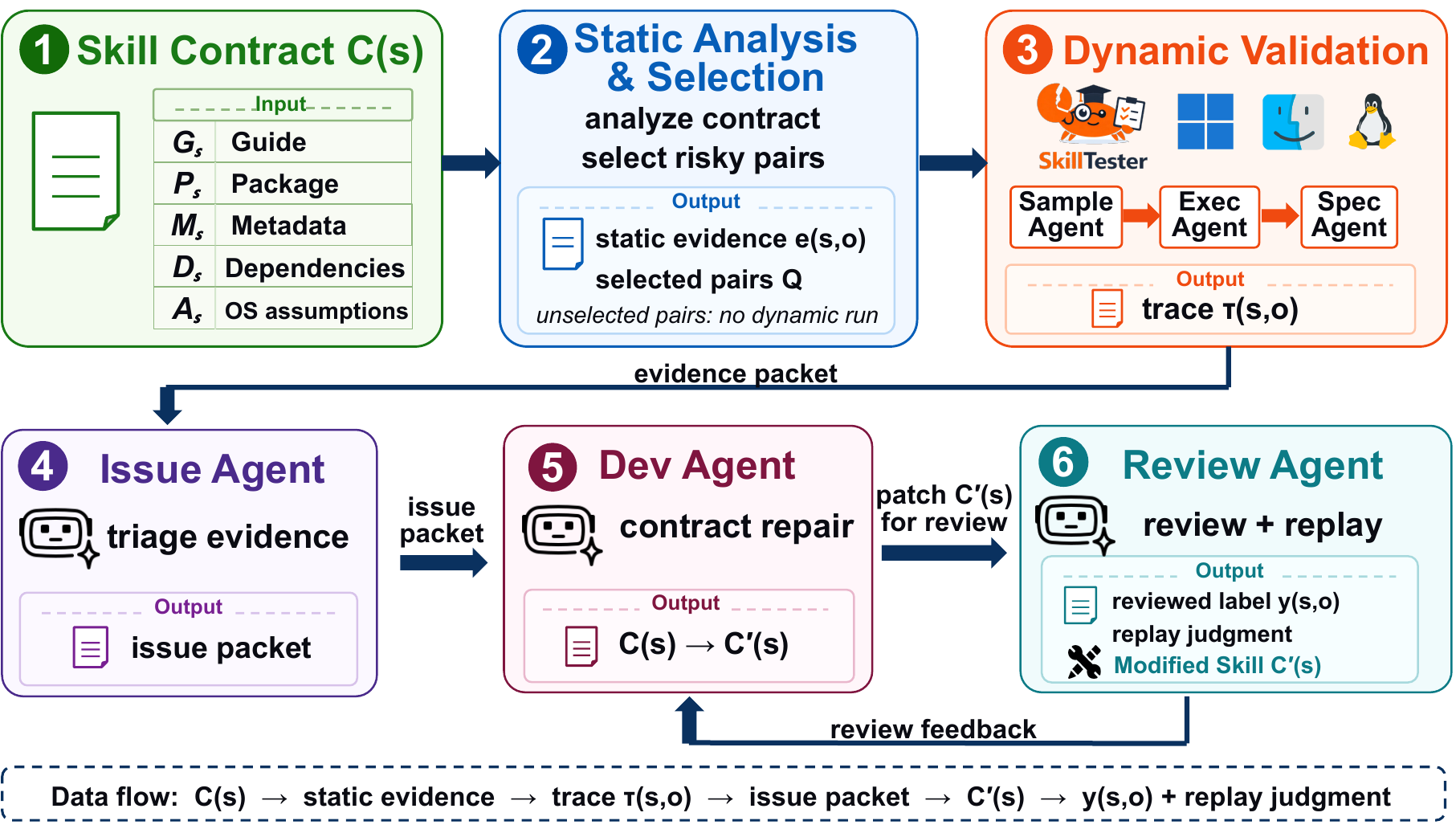}
  \caption{\tool workflow. Static evidence triages skill-OS pairs; review
  produces OS-root-cause labels; repair edits contracts and replays target
  tasks. SkillFab.ai~\cite{xu2026skillfab} provides the exchange substrate.}
  \label{fig:method-overview}
\end{figure*}

Figure~\ref{fig:method-overview} operationalizes the evidence chain as a
multi-agent workflow. The separation follows inspection and modern code
review~\cite{fagan1976inspection,sadowski2018review}: artifact generation and
repair are separated from root-cause labeling and replay judgment. Labels are
not inferred from inter-agent agreement. Each later stage consumes an explicit
packet and produces a scoped artifact, such as a dynamic trace, reviewed label,
contract edit, or replay judgment. The review role is a workflow role in
\tool, not maintainer acceptance; Section~\ref{sec:threats} treats its
agent implementation as a reliability threat.

\subsection{Design Invariants}

Three invariants keep \tool from confusing cheap evidence with a compatibility
label. Evidence precedes verdicts: static analysis over $C(s)$ is
recall-oriented and never assigns $y(s,o)$, because low-cost evidence can select
work but cannot by itself assign an OS-issue label. Outputs are contract-field
localized: each warning names the component that may be wrong, $G_s$, $P_s$,
$M_s$, $D_s$, or $A_s$, so the result can drive a concrete repair. Labels are
reviewed: a dynamic failure becomes a label only after root-cause review,
because credentials, services, task noise, and OS compatibility can produce the
same surface symptom.

Algorithm~\ref{alg:skill-checking} makes the budget boundary explicit. Static
analysis can only enqueue $(s,o,h)$ triples; \textsc{Select} chooses a
budgeted subset; \textsc{Review} creates labels; and \textsc{PlanRepair} runs
only after a confirmed, repairable label. The algorithm therefore returns
reviewed labels and a repair-action set rather than unreviewed static warnings.
This separates three actions that are often conflated in dynamic skill testing:
screening, labeling, and repair planning. \textsc{BuildContract} materializes
the contract fields defined in Section~\ref{sec:problem} from the skill guide,
package surface, metadata, dependencies, and OS assumptions.

\begin{algorithm}[t]
\caption{Contract diagnosis and repair}
\label{alg:skill-checking}
\begin{algorithmic}[1]
\Require skills $S$, target OS set $O$, static rules $\mathcal{R}$, dynamic budget $B$
\Ensure reviewed labels $Y$, repair-action set $\Pi$
\State $Y \gets \emptyset$; $\Pi \gets \emptyset$; $Q \gets \emptyset$
\ForAll{$s \in S$}
  \State $C(s) \gets \Call{BuildContract}{s}$
  \ForAll{$o \in O$}
    \State $(e(s,o),h(s,o)) \gets \Call{StaticAnalyze}{C(s),o,\mathcal{R}}$
    \If{$h(s,o) \ne \mathsf{skip}$}
      \State $Q \gets Q \cup \{(s,o,h(s,o))\}$
    \EndIf
  \EndFor
\EndFor
\ForAll{$(s,o,h) \in \Call{Select}{Q,B}$}
  \State $\tau \gets \Call{RunDynamicTasks}{s,o}$
  \State $y \gets \Call{Review}{h,\tau,C(s)}$
  \State $Y \gets Y \cup \{y\}$
  \If{$y$ is confirmed and repairable}
    \State $\Pi \gets \Pi \cup \{\Call{PlanRepair}{y,C(s)}\}$
  \EndIf
\EndFor
\State \Return $(Y,\Pi)$
\end{algorithmic}
\end{algorithm}

\subsection{Static Analysis}

Static analysis proposes which skill-OS pairs should enter the
dynamic-validation budget. For each pair $(s,o)$, the analyzer takes the contract
$C(s)$ and returns two objects: evidence $e(s,o)$ and a triage hypothesis
$h(s,o)$. It never assigns the final label $y(s,o)$, and
\textsc{Select} later enforces the budget $B$. Algorithm~\ref{alg:static-analysis}
gives the procedure used by Algorithm~\ref{alg:skill-checking}.

The analyzer evaluates a rule set $\mathcal{R}$ over the contract. Each rule
$\rho$ is attached to one contract component, such as $G_s$ or $P_s$, and is
interpreted through four predicates. \textsc{Match} extracts concrete
OS-sensitive snippets, \textsc{Reach} keeps only snippets used on the target-OS
execution or installation path, \textsc{Conflict} tests whether the snippet is
incompatible with the target OS, and \textsc{Fallback} suppresses snippets
already handled by a guarded branch, adapter, or explicit unsupported
declaration. A snippet fires only when it is matched, reachable, conflicting,
and not covered by a fallback.

\[
\begin{aligned}
\mathsf{fire}(\rho,C(s),o,u) \Leftrightarrow&
  \mathsf{match}_{\rho}(u,C(s)) \wedge
  \mathsf{reach}(u,o) \wedge\\
& \mathsf{conflict}_{\rho}(u,o) \wedge
  \neg\mathsf{fallback}(u,o,C(s)) .
\end{aligned}
\]

Algorithm~\ref{alg:static-analysis} stores each fired snippet as an evidence
record containing the implicated component, source location, rule, and snippet.
If no rule fires, the pair is skipped; otherwise, the pair enters dynamic
validation with ranked evidence. \textsc{Rank} prioritizes executable blockers
over wording mismatches, cross-field contradictions over isolated mentions,
missing fallbacks over guarded branches, and repairable contract defects over
platform-intrinsic requirements. For example, a \windows pair receives high
priority when $M_s$ claims Windows support, $P_s$ exposes only
\texttt{run.sh}, and $G_s$ instructs the agent to execute it.

\rev{We bootstrapped the initial rule set from compatibility features observed
in a seed set of 30 problematic skills. An LLM summarized these features into
34 candidate screening rules, which the authors then reviewed manually.}
Table~\ref{tab:static-rule-families} summarizes the six rule families, their
target-OS interpretations, and repair targets. These rules serve as screening
heuristics rather than independently validated predictors.
A positive $h(s,o)$ therefore means ``select this pair for dynamic
validation,'' not ``this is confirmed.'' Review still decides $y(s,o)$.

The analyzer does not simulate the agent, establish package portability, or
label a failure without execution evidence. It exposes low-cost contract
obligations before dynamic validation. Its main false-negative modes are
runtime-generated OS assumptions, live GUI or account-state failures, and
dependencies hidden behind remote services; these cases still require dynamic
validation and review.

\begin{algorithm}[t]
\caption{Static contract analysis for budget triage}
\label{alg:static-analysis}
\begin{algorithmic}[1]
\Require skill contract $C(s)$, target OS $o$, static rules $\mathcal{R}$
\Ensure evidence $e(s,o)$, hypothesis $h(s,o)$
\State $E \gets \emptyset$
\ForAll{$\rho \in \mathcal{R}$}
  \State $c \gets \Call{Field}{\rho}$
  \ForAll{$u=\langle\mathit{loc},\mathit{snippet}\rangle \in \Call{Match}{\rho,C(s)}$}
    \If{\Call{Fire}{$\rho,C(s),o,u$}}
      \State $E \gets E \cup \{\langle c,\mathit{loc},\rho,\mathit{snippet}\rangle\}$
    \EndIf
  \EndFor
\EndFor
\If{$E=\emptyset$}
  \State \Return $(\emptyset,\mathsf{skip})$
\EndIf
\State $E^\star \gets \Call{Rank}{E}$
\State \Return $(E,\mathsf{check}(E^\star))$
\end{algorithmic}
\end{algorithm}

\begin{table*}[t]
\centering
\caption{Static rule groups by skill-contract field.}
\label{tab:static-rule-families}
\begingroup
\scriptsize
\renewcommand{\arraystretch}{1.02}
\setlength{\tabcolsep}{3pt}
\begin{tabular}{@{}L{0.13\textwidth}L{0.27\textwidth}L{0.30\textwidth}L{0.24\textwidth}@{}}
\toprule
Contract field & Target-OS rule groups & Typical evidence & Repair target \\
\midrule
$G_s$ guide &
Shell, utility, install, and native-app command rules. &
\texttt{./run.sh}, \texttt{date -d}, \texttt{apt}/\texttt{brew},
\texttt{osascript}. &
Guarded guide alternatives or portable commands. \\
\midrule
$P_s$ package &
Missing entrypoint, single-OS wrapper, executable-bit, and native-binary rules. &
Absent referenced file, \texttt{*.sh}-only or \texttt{*.ps1}-only surface,
ELF/Mach-O/PE binary. &
Portable wrapper, missing file, or target-OS entrypoint. \\
\midrule
$M_s$ metadata &
Support-declaration and declared-runtime consistency rules. &
Advertised OS support, required binary/runtime, version or platform metadata. &
Correct support metadata or add the missing contract. \\
\midrule
$D_s$ dependency &
Package-manager, OS-specific CLI, and desktop/session dependency rules. &
\texttt{apt}/\texttt{brew}/\texttt{winget}, \texttt{xclip}/\texttt{pbcopy},
browser or accessibility requirement. &
Per-OS install recipe, dependency probe, adapter, or headless fallback. \\
\midrule
$A_s$ assumption &
Native API, filesystem namespace, path, permission, and GUI-assumption rules. &
\texttt{/proc}, AppleScript~\cite{appleScriptGuide}, COM~\cite{microsoftCOM},
\texttt{/tmp}, \texttt{chmod}. &
Adapter, runtime path API, guard, or unsupported declaration. \\
\midrule
Cross-field &
Contract-contradiction and fallback-recognition rules. &
Broad support claim with single-OS evidence; existing
\texttt{platform.system()} guard. &
Repair the authoritative field or suppress guarded evidence. \\
\bottomrule
\end{tabular}
\smallskip

\noindent\footnotesize\textit{Note:} Rules emit $e(s,o)$ and $h(s,o)$ after conflict, reachability, and
fallback checks.
\endgroup
\end{table*}

\subsection{Dynamic Validation}

Dynamic validation asks an agent to execute generated tasks for skill-OS pairs
selected by $h(s,o)$ under budget $B$. Dynamic execution is expensive evidence,
not the default first step for every pair. The trace $\tau(s,o)$
records whether the selected pair exhibits an execution symptom under the target
OS.

Each dynamic run records the target OS, selected task, agent transcript summary,
exit status, timeout, observed command errors, generated artifacts, and
available token and timing counters. The run record is broader than a
pass/fail bit: review needs enough context to distinguish an OS
compatibility defect from a missing account, unavailable service, benchmark
gap, or non-OS agent planning error.

\subsection{Root-Cause Review and Repair}

A failed dynamic run is only a symptom. \tool joins $e(s,o)$, $\tau(s,o)$,
package snippets, and target-OS context, then performs root-cause review.
\rev{Inspired by differential testing, the Review Agent compares task outcomes
and artifacts across operating systems when comparable executions of the same
task are available.}
Labels $y(s,o)$ distinguish confirmed OS compatibility defects from external
credentials, OAuth, API quota, private account state, benchmark capability
gaps, platform-intrinsic skills, non-OS task failures, and uncertain cases
requiring more evidence. This review step is necessary for precision and
defines the reference labels for static precision, recall, and F1. The review
packet for a skill-OS pair contains the static evidence record, the
relevant contract snippets, dynamic symptoms, and the review-stage decision.
The packet is the audit unit behind each table count: it links the contract
$C(s)$, static hypothesis $h(s,o)$, dynamic trace $\tau(s,o)$, and reviewed
label $y(s,o)$.

Confirmed and repairable labels are mapped back to the contract component that
explains the failure. Defects in $G_s$ receive OS-specific guide alternatives;
defects in $P_s$ receive portable wrappers or missing entrypoints; defects in
$M_s$ receive corrected support metadata; defects in $D_s$ receive target-OS
install contracts; and defects in $A_s$ receive adapters, fallbacks, permission
guards, or explicit unsupported declarations. The repair action $r$ transforms
$C(s)$ into $C'(s)$ and is evaluated using the target-OS replay protocol
defined in Section~\ref{sec:evaluation}.

\section{Evaluation}
\label{sec:evaluation}

The evaluation follows the evidence chain in Section~\ref{sec:problem}. RQ1
tests diagnosis and static triage, RQ2 tests contract-level repair, and RQ3
tests workflow value through repair trajectories, a direct-agent ablation, and
budgeted validation. The corpus contains 1,090 normalized skills, or 3,270
skill-target-OS cases. The
taxonomy contains 205 root-cause annotations because one case may expose more than one contract
defect. The budget analysis uses 1,298 audited \linux/\macos validation
candidates with a \linux-measured runner-resource proxy. Each result names its
empirical unit.

\subsection{Research Questions}

RQ1 asks whether failures can be localized to OS assumptions and whether static
analysis can triage them before dynamic validation. RQ2 asks whether reviewed
root causes identify useful contract repairs, measured by target-OS replay over
the same 175 cases. RQ3 asks whether the workflow matters beyond the endpoint:
repair trajectories, a direct-agent repair ablation, and policies for selecting
which pairs receive dynamic traces $\tau(s,o)$ under budget.

\subsection{Data Sources and Metrics}

\rev{We prioritized ranked entries from ClawHub and SkillSh and retained
1,090 normalized skills represented in all three OS datasets. Static triage
then selected skill--OS pairs for dynamic validation.}
Dynamic replays use OpenCode as the agent harness, so each run includes the
model, tool loop, file system, shell, and task
harness~\cite{opencode2026,anthropicHarnesses2025}.
Target-OS replays use three runner profiles: Ubuntu 22.04 for \linux, Apple
Silicon macOS for \macos, and Windows 11 for \windows.
\rev{The RQ1/RQ2 cohort comprises 175 skill--OS pairs with reviewed OS issues
and comparable target-task replays before and after repair. The cohort includes
cases with unchanged or lower post-repair pass counts.} Supporting units are the
205 root-cause annotations and the 1,298 audited \linux/\macos validation
candidates used only for budget analysis.

Each review packet contains the skill identifier, target OS, static evidence,
contract snippets, dynamic symptoms, exclusion evidence, and repairability
notes. Review assigns a target-OS symptom, links it to a violated contract
assumption, and identifies whether $G_s$, $P_s$, $M_s$, $D_s$, or $A_s$ is
repairable. Account, service, benchmark-gap, rate-limit, and non-OS planning
failures are excluded; ambiguous packets remain uncertain. ``Reviewer-judged''
therefore denotes this review-stage decision, not maintainer acceptance.

For static screening, positives are the pairs where $h(s,o)$ selects dynamic
validation, and reference positives are the pairs where
$y(s,o)=\mathsf{confirmed}$. Static precision, recall, and F1 are computed only
for pairs with both a static decision and a reviewed label, not over the full
corpus. We compute them as:

\[
\begin{aligned}
\mathit{precision} &= \frac{TP}{TP+FP},\\
\mathit{recall} &= \frac{TP}{TP+FN},\\
F1 &= \frac{2\cdot precision\cdot recall}{precision+recall}.
\end{aligned}
\]

\rev{Recall measures the fraction of reference-positive pairs selected for
dynamic validation, and precision measures the fraction of selected pairs that
are reference-positive. Appendices~\ref{app:static-screening} and
\ref{app:execution-replay} describe the screening evaluation and replay
protocol.} For efficiency analysis, we log
dynamic trace selection and, where the runner exposes them, non-cache tokens,
cache-read tokens, agent execution time, wall-clock time, and review metadata.

For RQ2, the baseline is the original, unrepaired skill contract $C(s)$. The
replay label comes from rerunning the same target-OS task that exposed the issue
after the contract change, with the review stage assigning the replay judgment.
Because the study does not consistently replay all
previously passing OSes, RQ2 reports target-OS repair effectiveness rather than
cross-OS repair correctness.

The direct-agent ablation uses the same 175 cases, issue packets, harness,
budget setting, and replay tasks as \tool, but removes the issue/submission
exchange and review-feedback loop. We report it in RQ3 because it isolates the
workflow contribution from the endpoint repair effect measured in RQ2.

For RQ3, full dynamic validation runs every validation candidate in the audited
sample. Static-guided validation runs only candidates selected by $h(s,o)$.
Uniform random selection samples validation candidates from the same
reviewed-label universe and estimates the expected issue yield. If that universe
has $N$ candidates and $m$ reviewed OS-issue candidates, a random budget of $b$
validations has expected yield:

\[
  \mathbb{E}[\mathit{issues}_{rand}(b)] = b\cdot \frac{m}{N}.
\]

Conversely, matching a target yield $z$ under uniform random selection requires
$zN/m$ validations in expectation. The two random policies test whether static
evidence increases issue yield per validation, while full dynamic validation
measures the audited-sample recall and execution effort.

\subsection{RQ1: Static and Dynamic Evidence for Cross-OS Skill Issues}
\label{subsec:rq1}

RQ1 tests the first step of the evidence chain: whether a cross-OS failure can
be explained by an OS assumption in the skill contract, and whether static
evidence is useful before running dynamic validation. The cohort is the
1,090-skill corpus described above. When multiple package versions represent the
same skill, the evaluation counts that skill once. The result table counts a
target-OS case only when review identifies at least one OS issue and the repair
study has comparable replay evidence for the original and repaired contracts.
The taxonomy then counts root-cause annotations, because one case can contain
multiple reviewed defects. Dynamic failures remain symptoms until review links
them to a concrete OS assumption in the contract.

\begin{table}[t]
\centering
\caption{RQ1 OS-related target-OS cases and static-triage performance.}
\label{tab:issue-characterization}
\footnotesize
\setlength{\tabcolsep}{2.4pt}
\begin{tabular*}{\columnwidth}{@{\extracolsep{\fill}}lrrrr@{}}
\toprule
Target OS & OS-related cases & Prec. & Rec. & F1 \\
\midrule
Linux & 42 & 0.276 & 0.640 & 0.386 \\
macOS & 25 & 0.173 & 0.688 & 0.277 \\
Windows & 108 & 0.242 & 0.719 & 0.362 \\
\midrule
Overall & 175 & 0.218 & 0.685 & 0.331 \\
\bottomrule
\end{tabular*}
\vspace{1pt}

Note: Counts use the 1,090-skill corpus per OS; P/R/F1 use records with both static decisions and reviewed labels.
\end{table}

The replay-aligned cohort reveals recurring OS-assumption patterns rather than
isolated anecdotes. Across the three OSes, 175 target-OS cases require an
OS-specific explanation and later repair replay. Since each OS is evaluated
over the same 1,090 skills, the per-OS confirmed-case rates are 3.85\% on \linux,
2.29\% on \macos, and 9.91\% on \windows. The distribution is uneven:
\windows contributes the largest share, consistent with skills that encode
Unix-like command, path, or entrypoint assumptions and only expose them when run
on native \windows.

The static columns should be read as a screening result. They evaluate
$h(s,o)$ against reviewed labels $y(s,o)$ only where both decisions exist, not as
full-corpus prevalence estimates. Overall precision is modest because the
static analyzer is not the judge of compatibility; its role is to put likely
OS-sensitive pairs into the dynamic-review queue. The more important screening
property is recall: static evidence captures a majority, roughly two thirds, of
confirmed OS issues while leaving the final label to dynamic execution and
review. For \windows, the static signals are recomputed from package source, but
the reviewed-label universe is a Windows source-evaluation subset rather than
the exact \linux/\macos static-evaluation universe; we therefore interpret the
pooled P/R/F1 values as a screening operating point rather than a prevalence
estimate.

\begin{table*}[t]
\centering
\caption{RQ1 root causes over 205 reviewed OS-issue annotations.}
\label{tab:issue-patterns}
\scriptsize
\setlength{\tabcolsep}{3pt}
\begin{tabular}{@{}L{0.23\textwidth}r l L{0.50\textwidth}@{}}
\toprule
Root cause & Count & OS split & What review observed \\
\midrule
\rev{Command/path/shell assumptions} & \rev{86 (42.0\%)} & \rev{11/18/57} &
\rev{POSIX shell~\cite{posix2024} on native \windows, BSD/GNU flag
mismatches~\cite{gnuCoreutils,freebsdDate}, hard-coded paths, missing portable
wrappers, or Windows path, encoding, locale, and path-length
conventions~\cite{microsoftWindowsNaming,microsoftMaxPath}.} \\
Native app/API assumptions & 65 (31.7\%) & 31/11/23 &
AppleScript~\cite{appleScriptGuide}, COM~\cite{microsoftCOM}, X11, browser
automation, desktop control, or platform-intrinsic apps. \\
Dependency/install contracts & 31 (15.1\%) & 16/6/9 &
Install steps or required tools exist for one package manager, local CLI,
runtime, desktop app, or host service but not the target OS. \\
Support metadata mismatch & 16 (7.8\%) & 0/0/16 &
Metadata or guide text advertises support that the package surface or
dependency contract does not implement. \\
Permission/session model & 7 (3.4\%) & 3/3/1 &
GUI display, accessibility, TTY, account state, or desktop session is required
but not declared or guarded. \\
\bottomrule
\end{tabular}
\smallskip

\noindent\footnotesize\textit{Note:} Shares are over 205 annotations; OS split is \linux/\macos/\windows.
\end{table*}

The root-cause distribution explains why the contract abstraction is useful.
The dominant categories are not opaque runtime crashes; they are assumptions
that leave traces in the skill contract before execution. Command, path, and
shell assumptions dominate \windows-oriented defects, while native application,
API, dependency, and session assumptions explain many \linux and \macos cases.
This pattern is what makes static evidence useful without making it sufficient:
the package often exposes the suspect assumption, but review is still needed to
decide whether that assumption explains the observed failure.

Answer to RQ1. Cross-OS failures become actionable only after review ties them
to contract assumptions. In the replay-aligned cohort, 175 target-OS cases
receive such labels, and static contract analysis captures a majority of
confirmed OS issues at a recall-oriented screening point. Static analysis therefore does not
replace execution; it selects where expensive dynamic evidence and root-cause
review should be spent.

\subsection{RQ2: Repair Patterns and Repair Effectiveness}
\label{subsec:rq2}

RQ2 starts where RQ1 ends. Once review localizes a failure to a contract
assumption, the repair question is whether that root cause also identifies the
field that should change. The unit remains a target-OS case rather than a
unique package: one repaired skill may target multiple OSes, and the right
contract edit can differ by target. RQ2 therefore compares reviewer-judged
target-OS replay before and after repair over the same 175 confirmed cases.

\begin{figure*}[t]
  \centering
  \ifreviewcopy
    \includegraphics[width=0.88\textwidth]{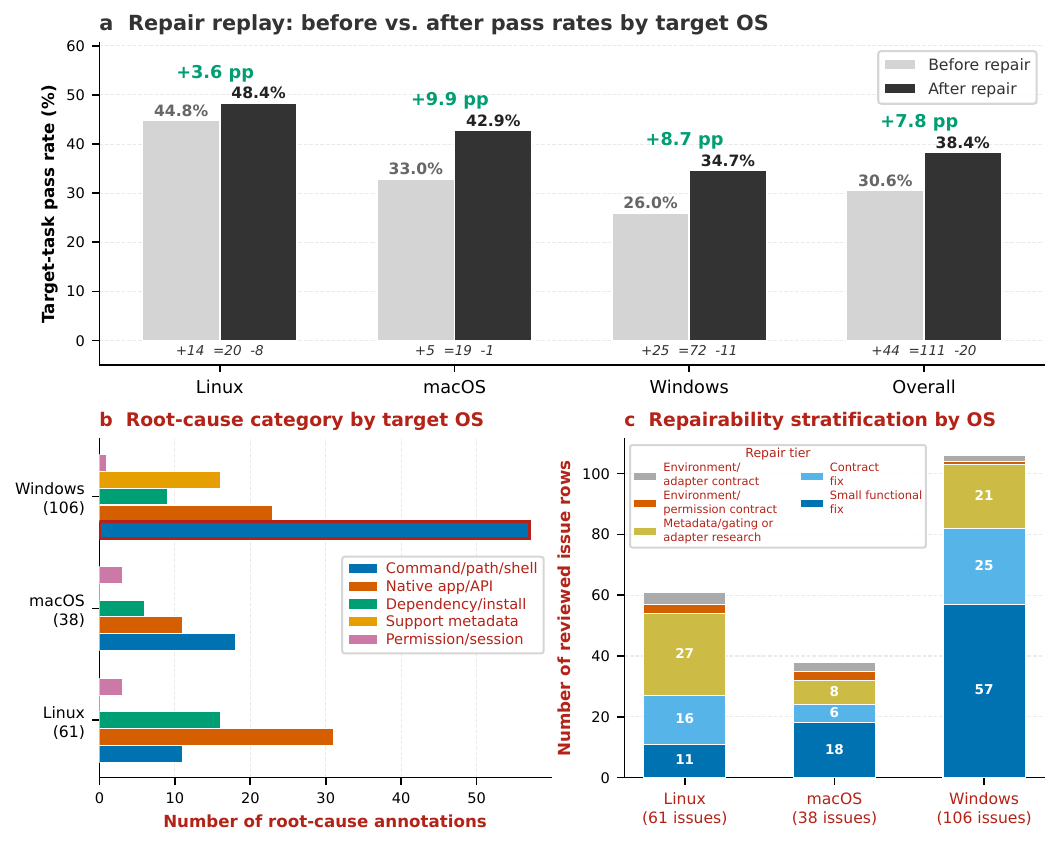}
  \else
    \includegraphics[width=0.88\textwidth]{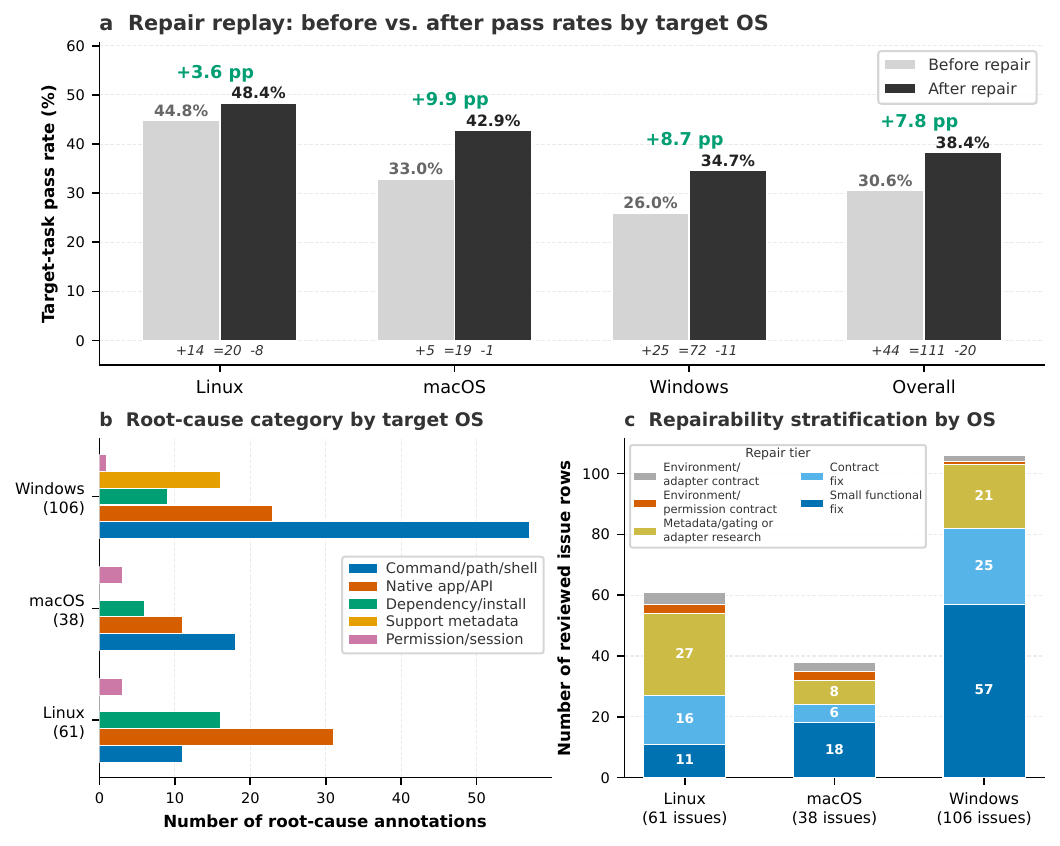}
  \fi
  \caption{RQ2 repair landscape: replay, root causes, and repairability.
    \rev{Panel (b) groups root-cause categories, whereas panel (c) groups
    repair tiers.}}
  \label{fig:repair-landscape}
\end{figure*}

Figure~\ref{fig:repair-landscape} summarizes the repair landscape. Its panels
separate replay outcomes, root-cause composition, and repairability rather than
mixing them into one denominator.
\rev{Repair tiers summarize the required repair action: small functional fixes,
contract fixes, metadata/gating or adapter research, environment/permission
contracts, and environment/adapter contracts.}

The mapping from diagnosis to repair is direct. Shell, command, and path
defects receive guarded command branches, portable wrappers, or missing
entrypoints; native API defects receive adapters, gates, or explicit unsupported
declarations; dependency defects receive per-OS install recipes or probes; and
support or session-model defects receive metadata and guard changes. The
effectiveness analysis includes skill-OS pairs whose repaired contract has the
original target-OS task set available before and after repair and a
reviewer-judged outcome for those tasks. This scope excludes intentionally
unsupported or platform-intrinsic cases from the target-replay denominator while
leaving them in the reviewed repair backlog.

The repair protocol has three constraints. A patch must address the reviewed
root cause rather than suppress the failing benchmark task; preserve existing
working OS branches or add explicit guards; and, for platform-intrinsic defects,
make the support boundary explicit rather than synthesize an unsupported
portability path. These constraints align RQ2 with contract preservation
instead of task-specific overfitting.

The replay study reuses the original benchmark tasks for each repaired case.
Figure~\ref{fig:repair-landscape} reports the target-OS repair effect over the
same 175 cases as RQ1. The aggregate pass rate rises from 30.6\% to 38.4\%.
At the case level, 44 cases improve, 111 remain unchanged, and 20 have lower
target-task pass counts after repair. These lower counts mark target-replay
risk, not regression on previously passing OSes.

\begin{figure}[t]
\centering
\ifreviewcopy\colorlet{listingrule}{diffred}\fi
\begin{lstlisting}[style=repairdiff]
scripts/office/soffice.py  (diff excerpt)
+SYSTEM = platform.system()
+_SOFFICE_BIN = "soffice.exe" if SYSTEM == "Windows" else "soffice"

 def get_soffice_env() -> dict:
     env = os.environ.copy()
-    env["SAL_USE_VCLPLUGIN"] = "svp"
-
-    if _needs_shim():
+    if SYSTEM in ("Linux", "Darwin"):
+        env["SAL_USE_VCLPLUGIN"] = "svp"
+
+    if SYSTEM == "Linux" and _needs_shim():
         shim = _ensure_shim()
         env["LD_PRELOAD"] = str(shim)
\end{lstlisting}
\caption{\rev{RQ2 repair of LibreOffice runtime configuration in
\texttt{xlsx-cn} (\texttt{SFOS-0150}). The repaired package passed 9 of 9
Windows replay tasks, up from 5 of 9 before repair.}}
\label{fig:rq2-repair-diff}
\end{figure}

\rev{Figure~\ref{fig:rq2-repair-diff} shows a repair that separates
platform-specific runtime settings from the shared spreadsheet recalculation
workflow. The original LibreOffice helper used a single executable name and
applied Unix-oriented settings without an OS check. The patch selects
\texttt{soffice.exe} on Windows, keeps the \texttt{svp} backend on Linux and
macOS, and limits the optional socket shim to Linux.} Other successful repairs follow the
same principle for missing entrypoints, dependency recipes, and support-contract
guards. The result packet records $e(s,o)$, the modified contract fields, a
diff-style patch summary, target-OS replay results, additional replay evidence
when present, and the review-stage judgment. The unit of repair is therefore
the contract field that caused the failure, not the individual benchmark task.

Answer to RQ2. Reviewed root causes can be converted into contract-level
repairs that improve reviewer-judged target-OS replay. Over the same 175
target-OS cases, the pass rate rises from 30.6\% for the original skills to
38.4\% after repair (+7.8\%).
The figure keeps case-level movement visible, so the result remains a target-OS
repair measurement rather than a full cross-OS correctness audit.

\subsection{RQ3: Repair Dynamics, Ablation, and Efficiency}
\label{subsec:rq3}

RQ3 asks what the workflow adds beyond the final repaired package. We separate
three effects: whether repair improves monotonically across rounds, whether the
issue/submission review loop changes outcomes relative to direct patching, and
whether static triage reduces the dynamic validation needed to find confirmed
OS issues.

\begin{figure}[!t]
\centering
\includegraphics[width=\columnwidth]{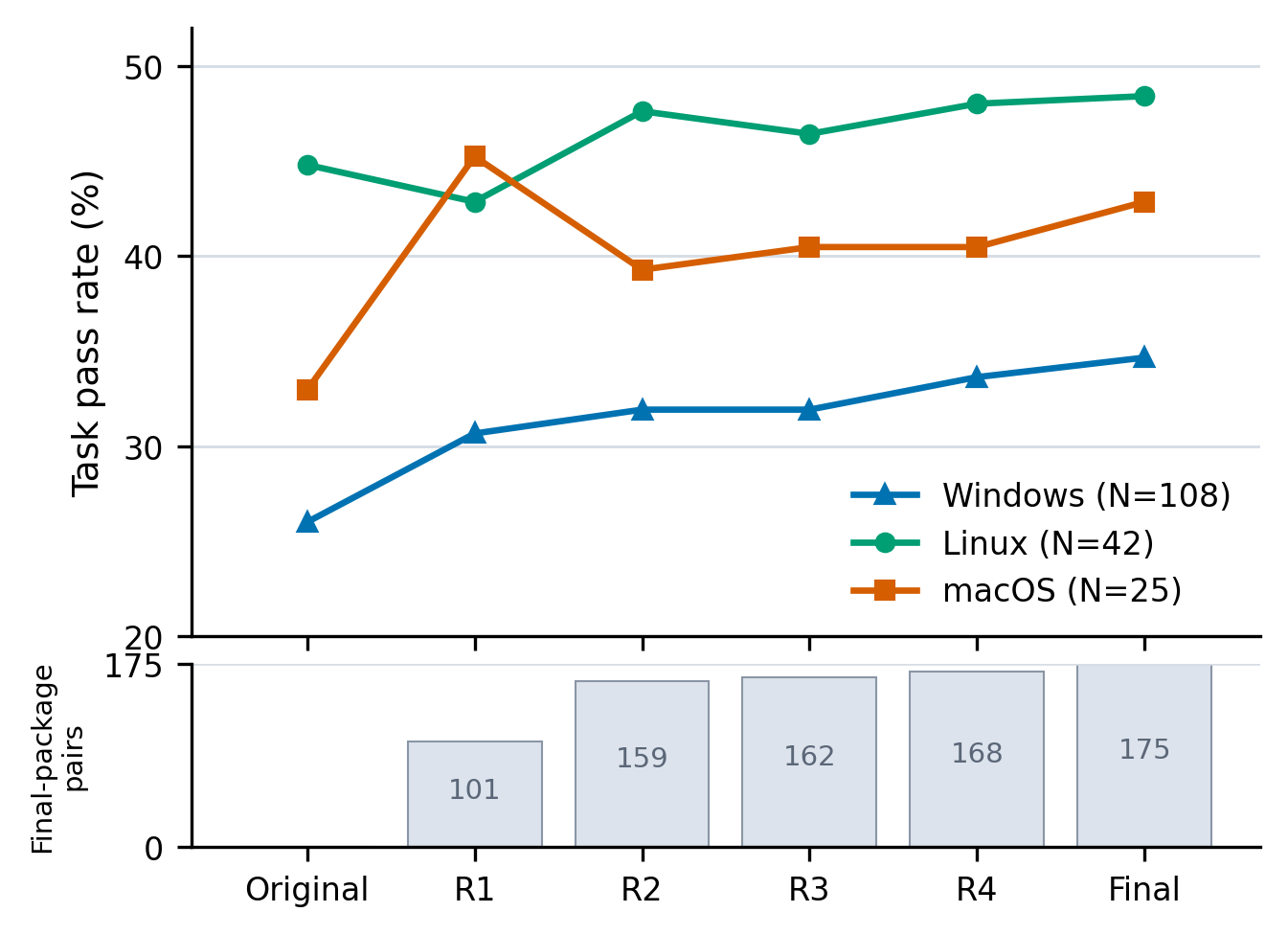}
\caption{RQ3 repair progression over 175 target-OS cases. Missing rounds use
carry-forward.}
\label{fig:rq3-round-progression}
\end{figure}

Figure~\ref{fig:rq3-round-progression} answers the first question. Repair is
not monotonic: an intermediate package can improve one target OS, expose a new
contract inconsistency, or fall below the final endpoint. The carry-forward rule
keeps the denominator fixed, so dips and plateaus reflect repair churn rather
than changing sample composition.

The direct-agent ablation uses the same 175 cases, issue packets, OpenCode
harness, budget setting, and replay tasks as \tool, but removes the
issue/submission exchange and review-feedback loop. The comparison is therefore
a workflow ablation rather than a change in model, task set, or replay
protocol.

\begin{table}[!t]
\centering
\caption{RQ3 direct-agent workflow ablation over 175 target-OS cases.}
\label{tab:rq3-direct-agent}
\scriptsize
\setlength{\tabcolsep}{2.0pt}
\begin{tabular*}{\columnwidth}{@{\extracolsep{\fill}}llrr@{}}
\toprule
OS & Method & Pass rate ($\Delta$) & $+{/}{=}{/}{-}$ \\
\midrule
Linux & Direct Agent & 46.8\% (+2.0) & 13/18/11 \\
Linux & \tool Workflow & 48.4\% (+3.6) & 14/20/8 \\
macOS & Direct Agent & 42.2\% (+9.3) & 8/16/1 \\
macOS & \tool Workflow & 42.9\% (+9.9) & 5/19/1 \\
Windows & Direct Agent & 33.7\% (+7.7) & 28/61/19 \\
Windows & \tool Workflow & 34.7\% (+8.7) & 25/72/11 \\
\midrule
All & Direct Agent & 37.4\% (+6.8) & 49/95/31 \\
All & \tool Workflow & 38.4\% (+7.8) & 44/111/20 \\
\bottomrule
\end{tabular*}
\vspace{1pt}

Note: $\Delta$ compares each method with \textsc{Original}; $+{/}{=}{/}{-}$ counts improved/unchanged/lower target-OS cases.
\end{table}

Table~\ref{tab:rq3-direct-agent} shows that Direct Agent recovers much of the
aggregate pass-rate gain, but \tool keeps the endpoint slightly higher and has
fewer lower target-replay outcomes (20 versus 31). The workflow benefit is
therefore not a large aggregate-pass-rate jump alone; it is a more controlled
path from issue evidence to patch and replay.

The third part of RQ3 treats static analysis as a selection policy.
Table~\ref{tab:rq3-efficiency} uses the audited \linux/\macos static-triage sample
because it contains both reviewed labels $y(s,o)$ and static decisions
$h(s,o)$. \textsc{Full dynamic} is a reference row that validates all 1,298
candidates. The controlled comparison fixes the target yield at 38 confirmed OS
issues and asks how many validations random selection would need to match
\tool.

The resource columns are estimated from a representative \linux runner trace
with complete token and time counters; the \macos and \windows replay releases
support outcome analysis but do not expose comparable runner token logs. The
cost model applies official DeepSeek-V4-Flash prices from the DeepSeek
website\footnote{\url{https://www.deepseek.com/}} to logged cache-miss
input, cache-hit input, and output tokens; it
excludes human review time and machine-rental cost.

\begin{table}[!t]
\centering
\caption{RQ3 static-triage efficiency on 1,298 reviewed \linux/\macos
candidates.}
\label{tab:rq3-efficiency}
\scriptsize
\setlength{\tabcolsep}{2.2pt}
\begin{tabular*}{\columnwidth}{@{\extracolsep{\fill}}lrrrrrrr@{}}
\toprule
Policy & Runs & OS issues & Recall & NC tok. & Cache tok. & Cost & Agent h \\
\midrule
Full dynamic & 1,298 & 57.0 & 1.000 & 391.1M & 3.95B & \$77.6 & 170.7 \\
\midrule
\multicolumn{8}{@{}l}{To find 38 OS issues} \\
Random w/o static & 865 & 38.0 & 0.667 & 260.7M & 2.63B & \$51.7 & 113.8 \\
\tool & 185 & 38.0 & 0.667 & 55.7M & 562.9M & \$11.1 & 24.3 \\
\bottomrule
\end{tabular*}
\smallskip

\noindent\footnotesize\textit{Note:} Random w/o static is uniform random selection from the same reviewed
candidate set. NC tok. denotes non-cache tokens. Costs use \linux runner
counters and DeepSeek-V4-Flash pricing from the DeepSeek website.
\end{table}

This is a selection-efficiency result. To reach the same expected yield of 38
confirmed OS issues without static triage, random selection would require about
865 validations. \tool reaches that yield with 185 validations. This is a
\linux/\macos yield-per-validation result, not a new three-OS prevalence
estimate.

\vspace{0.5\baselineskip}
Answer to RQ3. \tool's value is process control and allocation efficiency, not
only a final pass-rate gain. Intermediate repairs churn, Direct Agent reaches
much of the aggregate improvement but produces more lower replay outcomes, and
static triage concentrates validation on higher-yield target-OS cases. In the
audited \linux/\macos validation-candidate sample, matching \tool's issue yield
by uniform random selection would require about 680 additional validations,
205.0M additional non-cache runner tokens, 2.07B additional cache-read tokens,
\$40.6 additional official-model cost, and 89.5 additional agent-hours under
the \linux-measured resource proxy.

\section{Discussion}

The results support one design principle: cross-OS skill compatibility should
be handled as a contract-linked evidence chain, not as a task pass/fail bit.
Static analysis is therefore an allocation signal rather than a final judge. It
uses cheap contract evidence to decide where dynamic validation is worth
spending; in the audited validation-candidate sample, matching \tool's 38
confirmed OS issues by uniform random selection would require about 865
validations rather than 185.

This framing also explains the precision/recall tradeoff in RQ1. A static
warning is not meant to be a compatibility verdict, so modest precision mainly
increases review load. Missed OS issues are more costly because they never
receive dynamic evidence under a fixed budget. \tool therefore treats static
analysis as a recall-oriented queueing mechanism and leaves confirmation to
execution and review.

The skill contract is the unit of explanation and repair. Static analysis
extracts evidence from $C(s)$, dynamic execution records symptoms, review maps
symptoms to $y(s,o)$, and repair transforms $C(s)$ into $C'(s)$. This chain
keeps guide text, package files, dependencies, and OS assumptions tied to one
object, which is necessary because \windows defects often involve Unix-style
commands and paths, \macos defects involve Bash, BSD/GNU tools, sessions, and
permissions, and \linux defects involve native-app, GUI, session, and dependency
gaps.

Repair is a workflow property rather than a single patch property. RQ2 therefore
reports reviewer-judged target-OS replay instead of package-wide cross-OS
correctness. The direct-agent baseline shows that a strong agent can recover
much of the aggregate pass-rate gain, but \tool keeps the endpoint slightly
higher and has fewer target-OS cases in the $-$ column. RQ3 then shows why the
process matters: intermediate packages can churn, so patch generation, review,
and replay should remain separate artifacts.

The baseline result should not be read as evidence that direct agents cannot
repair skills. It shows a narrower point: when repairs change mixed artifacts,
the issue packet, patch, review judgment, and replay record need to remain
linked. That traceability is what turns a passing patch attempt into a repair
claim that another reviewer can inspect.

The same lesson applies to skill repositories. A registry that stores only a
skill package and a task pass rate cannot explain whether a later failure is
caused by the guide, package surface, metadata, dependencies, or OS assumptions.
Skill repositories should therefore preserve contract-level evidence: declared
platform support, referenced entrypoints, install recipes, known native APIs,
and replay summaries. These records are lightweight compared with full dynamic
execution, but they give later portability review and repair concrete evidence
to inspect.

The results also suggest a benchmark design principle. Cross-OS skill benchmarks
should report not only whether an agent completed a task, but also which
contract assumption the run exercised and which OS boundary the task crossed.
Otherwise, a benchmark mixes OS defects, service failures, missing credentials,
and planning noise into one failure column. \tool does not eliminate this
ambiguity automatically; it gives the workflow a place to record and review it.

For deployment, this argues for a tiered checking policy. A skill registry can
run static contract checks whenever a package changes, flag likely target-OS
conflicts, and reserve dynamic validation for pairs whose evidence is reachable
and unguarded. This policy does not require every submitted skill to be executed
on every OS before review, but it still gives reviewers concrete evidence when a
skill claims broad platform support.

The multi-agent design follows from the same constraint. Generation, execution,
root-cause review, and repair are different software-engineering tasks, and
collapsing them into one prompt hides the evidence boundary. Separating the
roles makes the workflow slower than a single direct patch attempt, but it
creates the artifacts that make a repair claim auditable: the static record, the
dynamic trace, the reviewed label, the patch, and the replay result.

\section{Threats to Validity}
\label{sec:threats}

\subsection{Construct Validity}

Skill compatibility spans guide text, package
layout, dependencies, and runtime behavior, so a task failure is only a symptom.
\tool counts an OS issue only when review links the symptom to an OS assumption
in the skill contract. \rev{The measured repair effect may not generalize to
OS issues for which no replay evidence is available.}
The contract abstraction may also miss defects visible only through private
service state, runtime-generated behavior, or an agent workaround.

\subsection{Internal Validity}

Dynamic outcomes depend on the agent,
task generator, runner profile, timeout, credentials, and OS state. RQ2 therefore
reports reviewer-judged target-OS replay rather than full cross-OS repair
correctness; the study does not consistently replay every previously passing
task on every OS. The direct-agent baseline is retrospective over the same issue
packets and replay tasks, so it is a workflow ablation rather than a randomized
study over newly sampled repairs.

The review role is agent-implemented, so labels and replay judgments may differ
from human maintainer acceptance. \tool reduces this risk by storing review
packets with static evidence, contract snippets, dynamic symptoms, modified
fields, and replay outcomes for later adjudication.

\subsection{External Validity}

The corpus comes from ClawHub and
SkillSh, so results may differ for private enterprise skills, later registries,
or long-tail skills outside the ranked sample. The three target OSes cover
common desktop and server environments, not mobile, container-only, or embedded
systems. The observed root-cause mix may also shift as skill repositories add
platform metadata, per-OS install recipes, or CI checks.

\subsection{Reliability and Reproducibility}

The RQ3 cost comparison is retrospective over reviewed \linux/\macos validation
candidates, expected random-selection baselines, and a \linux-measured
runner-resource proxy. It estimates allocation value and token/time savings
rather than reporting additional randomized executions or directly measured
\macos/\windows token costs. Reproducibility is bounded by credentials and
private execution state; the artifact therefore releases summary data, code,
labels, and redacted symptom records rather than secrets or raw private traces.

\section{Related Work}

\subsection{Portability and Static-Guided Validation}

Software portability and empirical porting studies motivate designing and
measuring software across environments~\cite{mooney1990,iso25010,
ray2012crosssystem,pappas2024macro,silva2026portability}. Static-guided testing
and industrial static-analysis experience show that cheap structural evidence is
most useful when it guides expensive validation rather than serving as the final
judge~\cite{godefroid2005dart,sadowski2018static}. \tool applies this
cost-aware view to skill-OS pairs, whose contract spans guide text, package
files, metadata, dependencies, entrypoints, and OS assumptions.
The distinction is that portability is not only a property of executable source
code here. A skill can be non-portable because its prose names the wrong
command, its metadata overstates support, or its install contract omits a target
OS. This is why \tool treats static analysis as contract triage rather than as a
source-only portability checker.

\subsection{Agent Skills and Benchmarks}

Agent and multi-agent systems show how LLM agents invoke external capabilities,
coordinate software tasks, and package reusable skills~\cite{chen2021codex,
qian2024chatdev,hong2024metagpt,xu2026skillfab}. SkillTester-style evaluation
makes third-party skill behavior observable through generated tasks~\cite{
wang2026skilltester}, but a pass/fail trace does not explain whether an OS
assumption, account state, service dependency, benchmark gap, or agent planning
error caused the outcome. \tool adds the missing contract and reviewed-label
layer between dynamic traces and compatibility claims.
This makes the contribution complementary to skill platforms and skill
benchmarks. Platforms provide the place where skills, issues, and patches can be
stored; benchmarks provide tasks and executions. \tool specifies how to connect
those records into an OS-compatibility diagnosis and a repair claim.

\subsection{Repair and Review Workflows}

Automated repair and code-review work motivates reporting repair yield with
validation cost and workflow constraints~\cite{legoues2012genprog,
marginean2019sapfix,repairagent2025,fagan1976inspection,sadowski2018review}.
\tool uses that discipline for mixed skill contracts: repairs may change prose,
scripts, metadata, dependency contracts, or OS guards, and the claim remains
reviewed target-OS replay over $C(s)\rightarrow C'(s)$ rather than proof of
package-wide cross-OS correctness.
The repair target is therefore broader than a program patch but narrower than
free-form prompt editing. A reviewed label names the contract component that
failed, and the repair is expected to change that component while preserving the
skill's stated intent.

\section{Conclusion}

Cross-OS skill compatibility is a software-engineering problem for mixed
text-code-package artifacts. \tool links static evidence, dynamic symptoms,
reviewed root causes, and repair actions in one contract abstraction, then uses a
review-separated workflow to produce scoped repairs. \rev{The reviewer-judged
target-task pass rate rises from 30.6\% to 38.4\% after repair.} In the audited
\linux/\macos budget analysis, static-guided selection reaches the same expected
issue yield with 185 validations rather than about 865 under uniform random
selection. The repair claim remains scoped to reviewer-judged target-OS replay
rather than package-wide cross-OS correctness.

\section{Data Availability}
\label{sec:data-availability}

Our experimental code and data are available through the public SkillBridge
repository at \rev{\url{https://github.com/TankTechnology/SkillBridge}}.

\appendices
\section{\rev{Static-Screening Evaluation}}
\label{app:static-screening}
\rev{We evaluated static screening at the skill--target-OS level. We matched
each screening decision to its root-cause review label and included both flagged
and unflagged pairs with definitive labels within the evaluation scope. A pair
counted as a predicted positive if the static analyzer selected it for dynamic
validation. It counted as a reference positive if the Review Agent confirmed an
OS-related defect. We computed precision, recall, and F1 from these binary
decisions. Root-cause categories guided diagnosis and repair.}

\section{\rev{Execution, Root-Cause Review, and Repair Replay}}
\label{app:execution-replay}
\rev{Automated scripts drive agent harnesses, including OpenCode and Kimi Code,
using our task and review prompts. Root-cause review batches used
DeepSeek-V4-Flash. The review prompts request evidence references, OS
attribution, exclusions, and uncertainties. We retain the review inputs and
reports. The artifact includes the repair coordinator, repair prompts, and
configurable runner parameters; complete historical runner and external
patch-review deployment settings are outside this release slice.}

\rev{For each selected skill--OS pair, the execution agent runs the tasks on the
target OS according to their specifications and acceptance criteria. The
execution records capture task outcomes, generated artifacts, and relevant
command errors. Given these records and the relevant fields of the skill
contract, the Review Agent assesses whether the observed failure can be
attributed to an OS assumption. The Review Agent sees static predictions and
evidence as diagnostic hypotheses. Confirmation requires execution evidence
supporting an OS-related cause. Cases with insufficient evidence remain
uncertain.}

\rev{Repair agents modify implicated contract fields and receive patch-review
feedback. A repair is valid only when it addresses the reviewed field, preserves
the original task requirements, and is evaluated by target-OS replay. This
operationalizes repair effectiveness rather than package-wide semantic
correctness. Task pass rates aggregate replay outcomes, retaining comparable
cases with unchanged or worse outcomes.}

\section{\rev{Agent Roles and Operational Boundaries}}
\label{app:agent-roles}
\rev{Figure~\ref{fig:method-overview} names three SkillBridge agent roles. The
Issue Agent triages the evidence packet into a scoped issue packet. The Dev Agent
applies the contract repair and returns a patch for review. The Review Agent
performs review and replay, producing the reviewed label $y(s,o)$ and replay
judgment. Static Analysis \& Selection remains a preceding screening stage: it
selects risky pairs but does not assign OS labels. These are workflow boundaries
rather than a claim that every historical run used one fixed model or harness;
the documented execution and review configurations are stage-specific.}

\section*{\rev{Acknowledgments}}
\rev{This work was supported in part by the National Key Research and Development
Program of China under Grant 2023YFB3308500.}

\IEEEtriggeratref{30}
\bibliographystyle{IEEEtran}
\bibliography{references}

\end{document}